\documentclass[12pt,preprint]{aastex}

\newcommand{\etal}{et~al.}

\slugcomment{Accepted for publication in ApJ}

\shorttitle{Ages of Star Clusters in M33}
\shortauthors{Park, Park, and Lee}

\begin{document}


\title{Ages of M33 Star Clusters Based on the HST/WFPC2 Photometry}

\author{Won-Kee Park, Hong Soo Park, and Myung Gyoon Lee}

\affil{Astronomy Program, Department of Physics and Astronomy,
Seoul National University, Seoul 151-742, Korea}
\email{wkpark@astro.snu.ac.kr, hspark@astro.snu.ac.kr, mglee@astro.snu.ac.kr}

\begin{abstract}
We present a result of age estimation for star clusters in M33.
We obtain color-magnitude diagrams (CMDs) of
resolved stars in 242 star clusters from the HST/WFPC2 images.
We estimate ages of 100 
star clusters among these, by fitting the Padova 
theoretical isochrones to the observational CMDs. 
Age distribution of the star clusters shows a dominant peak at $log(t) \sim 7.8$. 
Majority of star clusters are younger than $log(t) = 9.0$, while ten star clusters
are older than $log(t) \sim 9.0$.
There are few clusters younger than $log(t) = 7$ in this study, which is in contrast
with the results based on the integrated photometry of star clusters in the previous studies.
Radial distribution of the cluster ages shows that
young to intermediate-age clusters are found from the center to the outer region,
while old clusters are distributed farther from M33 center.
We discuss briefly the implication of the results with regard to the formation of M33 
cluster system.
\end{abstract}

\keywords{galaxies: star clusters --- galaxies: individual (M33) ---
galaxies: photometry}

\section{Introduction}

Ages of star clusters in a galaxy are important information for understanding the 
formation and evolution of star clusters and their host galaxies.
In general we estimate ages of star clusters whose stars are resolved by fitting the
color-magnitude diagrams (CMDs) of the stars with theoretical isochrones, and 
ages of unresolved clusters by fitting their integrated photometry or spectra
with population synthesis models. 
Star clusters in the Local Group galaxies are partially resolved into individual stars
in the images obtained with the Hubble Space Telescope (HST)
so that they play an important role in the study of extragalactic star clusters.

There have been numerous studies that estimated the ages of star clusters in the Local
Group galaxies with the photometry of resolved stars in star clusters: for example,
LMC (e.g. \citet{mac06} and references therein) and SMC \citep{gla08},
NGC 185, NGC 205 \citep{sha06}, M31 \citep{ric05},  and M33 \citep{sar00, sar07a}.

M33 (NGC 598), Triangulum Galaxy, is a late-spiral galaxy (Scd), and is one of the
three spiral galaxies in the Local Group. It is known to have 
numerous star clusters (see \citet{sar07b, par07, zlo08} and references therein.).
It is located at about 900 kpc from us \citep{gal04} 
and has a large number of
star clusters so that it offers us an excellent opportunity to study 
the ages of resolved star clusters. 
At the distance of M33, bright star clusters are
easily resolved into stars and even the faint star clusters can be seen with a
hint of resolved stars on the images taken with HST/Wide Field and Planetary
Camera 2 (WFPC2).

There were several studies on the ages of star clusters in M33.
\citet{cha99b, cha01}  estimated the ages of M33 star clusters 
in the catalog of \citet{cha99a, cha01} by
comparing the integrated photometry of the star clusters with a simple
stellar population (SSP) model in the color-color diagram.
Later \citet{cha02} supplemented their age estimates for more M33 star clusters with
the spectroscopic data. 
They found that the 
age distribution of star clusters in the halo of M33 shows a large age spread,
and little progression of abundance with age. They pointed out these properties are
consistent with accretion origin for a large fraction of M33 halo but not with monolithic
collapse scenario of the halo.
From the comparison of globular cluster properties in nearby dwarf galaxies, they claimed M33 
halo cluster population might have formed by accretion of lower luminosity dwarf 
irregular galaxies and dwarf spheroidals rather the ones such as Magellanic clouds. 
The apparent lack of metallicity gradient among the ten
M33 halo cluster samples of \citet{sar00} also further supported the accretion scenario. 

\citeauthor*{ma01}
estimated the ages and metallicity of the M33 star clusters in their series of 
papers \citep{ma01, ma02a, ma02b, ma02c, ma04a, ma04b}.  
They obtained the integrated spectral energy distribution (SED) of star clusters from 
the photometry using the Beijing-Arizona-Taipei-Connecticut (BATC) system with 
13 narrow band filters. They fitted the SEDs with the SSP model to simultaneously 
determine the age and metallicity of star clusters. 
They showed that their age estimations for star
cluster samples from \citet{cha99a, cha01} are consistent with previous estimates.
Using the age estimates of M33 clusters given by \citeauthor*{ma01},
\citet{sar07b} showed that the number of clusters in the age distribution
appears to decline with age with no obvious breaks or abrupt changes, being
fit well by a single power law.

\citet{sar00} estimated ages of ten resolved globular clusters
in M33 using the isochrone fitting technique to the deep CMDs of resolved stars
derived from the HST/WFPC2 images.
They estimated the metallicity from the slope of red giant branch, and found that
M33 star clusters show a significant second parameter effect. They concluded from
this result that these star clusters might be several Gyrs younger than the Galactic
globular clusters if the second parameter is the age.
\citet{cha06} estimated the age of a globular cluster, C38 in the list of \citet{cs82},
to be $\sim$2-5 Gyrs old, utilizing the integrated spectrum and the resolved member 
star CMD.
\citet{sar07a} found 24 star clusters from the deep
images of two HST/Advanced Camera for Survey (ACS) fields in M33, and derived 
ages for 20 of them by fitting theoretical isochrones to the features on the CMDs 
of resolved stars including the main-sequence turnoffs (MSTO).
The derived ages of the star clusters range from $log(t) = 7.5$ to $log(t) = 10.1$.
Recently \citet{sto08} derived an age of $t >7$ Gyr,  for one extended cluster 
at the outer edge of M33, using the CMD of the resolved stars. \citet{zlo09} also
estimated ages of four M33 star clusters by fitting isochrones to the member star
CMD derived from HST/ACS images.

\citet{par07} compiled a comprehensive catalog of 242 star clusters in M33
identified from the HST image data. Along with 32 new star clusters detected
from 24 HST/WFPC2 fields, they combined several previous M33 star catalogs \citep{bed05,
cha99a, cha01, sar07a} obtained with HST data only, thus to ensure that the resulting
master catalog contains only the confirmed star clusters. This catalog offers a good 
sample of star clusters for age studies. 

As part of our project on studying star clusters in M33, we derived photometry
of the resolved stars in and around the star clusters from the HST/WFPC2 images
that contain M33 star clusters listed in \citet{par07}. 
We estimated ages of 100 star clusters in M33 using the resulting CMDs of
resolved stars.

Section 2 describes the HST/WFPC2 data used in this study, and the data reduction
procedures. Section 3 present the CMDs of M33 star clusters, and the estimates
of cluster ages.
In Section 4, we describe the age distribution of M33 star
clusters thus derived, together with their integrated photometric 
parameters and their radial distributions from the center of M33. 
In Section 5 we discuss differences
among the age estimations for M33 star clusters, and the implications of the results
with regard to the formation of M33 cluster system. Finally, summary and conclusion
are given in Section 6.

Throughout this paper, we will refer to the star clusters according to IDs
in the catalog of \citet{par07}. For cross-identifications, 
we will refer to the star clusters from the catalog of
\citet{bed05} with B designation, and ones from \citet{cha99a, cha01} with CBF.
CS designation means the star clusters from \citet{cs82}, MKKSS designation
means the ones from \citet{moc98}, MD the ones from \citet{md78}, and SBGHS
designation means the star clusters from \citet{sar07a}. 
Note \citet{sar07b} 
compiled all published lists of M33 star clusters to build a comprehensive catalog
which is being maintained on-line\footnote{
\url{http://www.astro.ufl.edu/\~{}ata/cgi-bin/m33\_cluster\_catalog/index.cgi}}. 
For reference, we included their catalog IDs  with SM designation.

\section{Data \& Data Reduction}

We used  HST/WFPC2 images of M33 fields available in the HST archive. 
These HST/WFPC2 image sets include the 20 fields
listed in Table~1 of \citet{cha99a}, 35 fields listed in Table~1 of \citet{cha01},
and 24 fields listed in Table~1 in \citet{par07}.

\begin{figure}
\epsscale{0.75}
\includegraphics[width=12cm, angle=270]{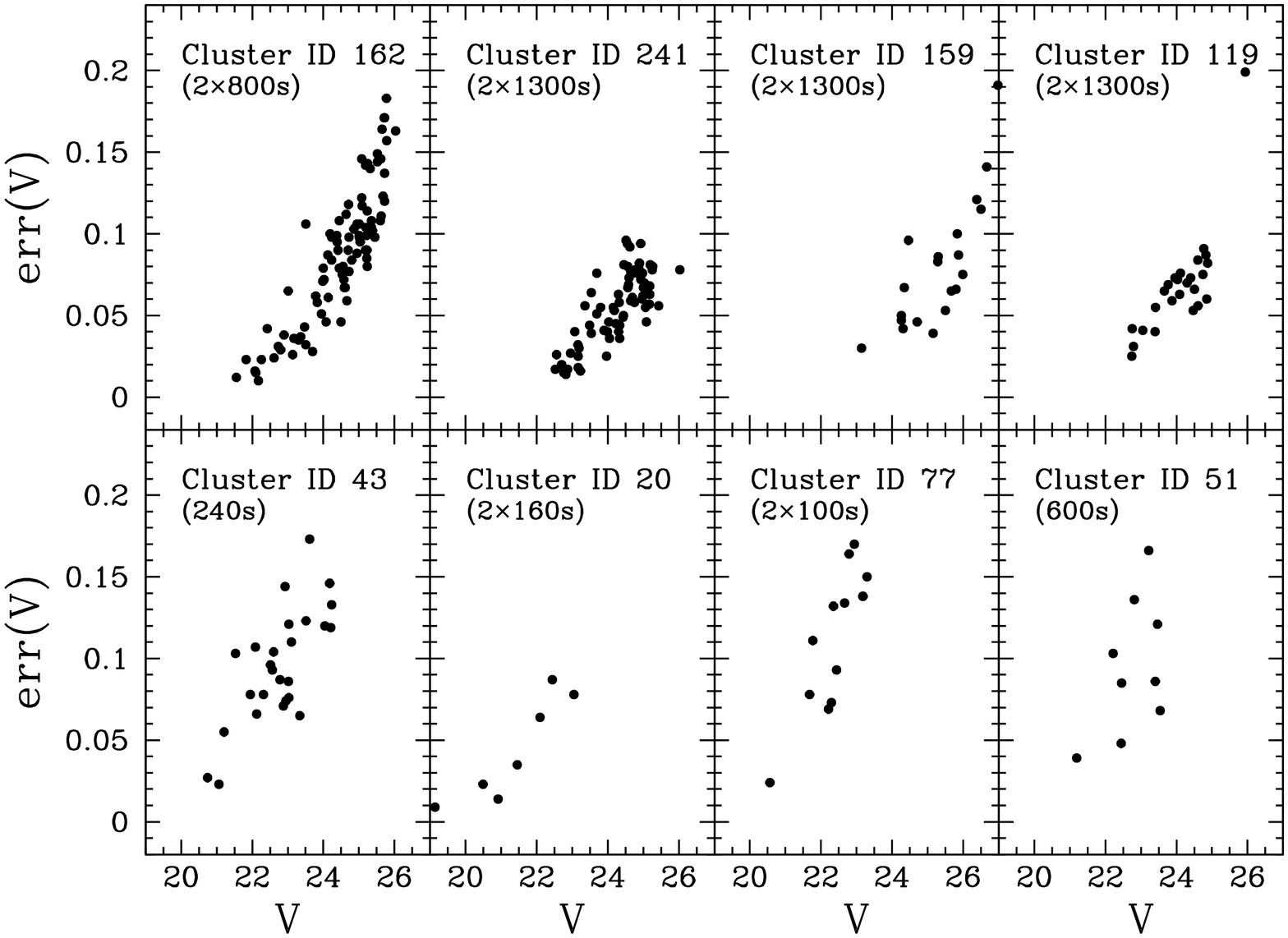} 
\caption{ 
Photometric errors versus $V$ magnitudes for a 
representative set of star clusters  in this study. 
Upper four panels show the photometry with exposures longer than
$1000$ seconds while lower four panels show the results with
exposures shorter than $1000$ seconds.
\label{magerr}}
\end{figure}

We derived the magnitudes of the point sources in the images using HSTphot 
\citep{dol00a} that is designed for the point spread function (PSF) fitting 
photometry of point sources in the WFPC2 images. 
HSTphot uses a library of TinyTim PSF for PSF fitting. Individual images
were all coadded with utility program included in HSTphot.
HSTphot produces instrumental magnitudes as well as standard magnitudes 
calibrated with \citet{hol95}'s relation and zeropoints from \citet{dol00b}
after it automatically applies the correction for Charge Transfer Efficiency
(CTE) \citep{dol00b}. Since HSTphot only calibrates the 
instrumental magnitudes of widely used filters, we used only the images obtained 
with 
$F436W$(B), $F555W$(V), $F606W$(V), and $F814W$(I).
The resultant photometry data turn out to have small photometric error. 
Fig.~\ref{magerr} shows that resultant data have photometric errors smaller
than $0.1$ at $V\sim 22$ mag even for about 200 sec long image, and
as for the data from exposures longer than 1000 sec, the errors get
larger than $ 0.1$ only for $V > 24$.

\section{Results}

\subsection{Age Estimation}

Apparent sizes of the star clusters in the images are $r \sim 1.0 \arcsec$ in most cases, 
and as large as $r \sim 3\arcsec$ for some cases with long exposures.
We construct the CMDs of resolved stars in cluster-centric radii from $1\arcsec$ to
$2.5\arcsec$
varying on the sizes of the clusters, which are 
listed in the fifth column of Table~\ref{cluster_age}.
To minimize the field star contamination, the CMDs of field stars were compared
with those of member stars around each star cluster. Field stars were selected
in the annulus of which the inner radius is $r \sim 3\arcsec$ and 
of the same area as the member star selection circle.

Since HST/WFPC2 data used in this study were obtained
from various observational projects, the depths of the images are not homogeneous.
Thus, it happened that a star cluster observed on more than one field in our image sets
resulted in different individual star detection on each appearance. In such a case, we 
used the CMD that has the largest number of detected stars for the analysis.

\begin{figure}
\epsscale{0.90}
\plotone{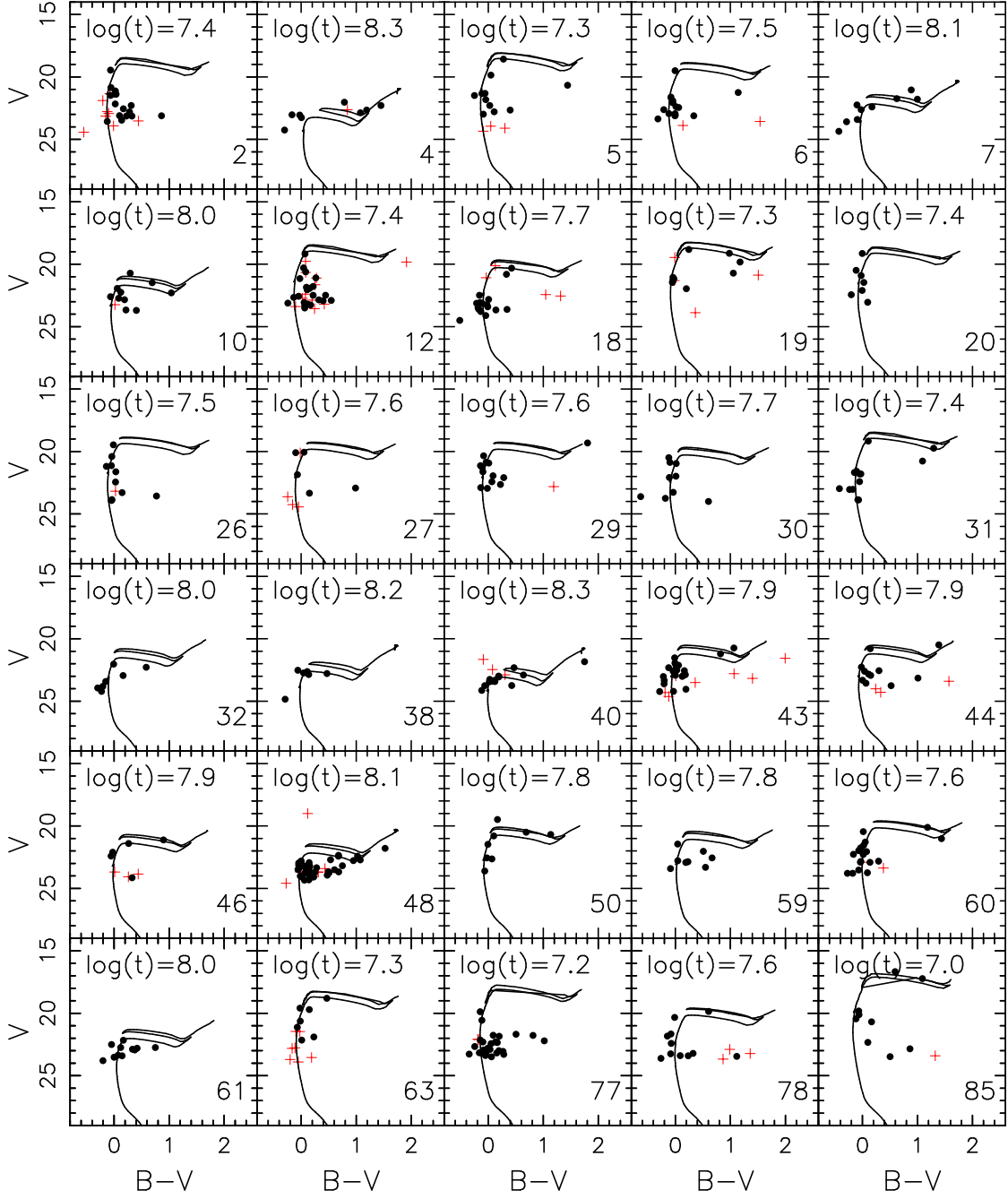} 
\caption{ 
$V-(B-V)$ CMDs of the resolved stars in M33 star clusters. 
Filled circles denote the stars found within the 
member selection radius of each star cluster, and crosses denote
the field stars around the star clusters.
Solid lines represent the Padova theoretical isochrones fit to the data. 
ID and derived age of each star cluster are shown at the lower right 
corner and upper left corner of each plot, respectively. 
\label{starcmd}}
\end{figure}

Fig.~\ref{starcmd} displays the CMDs of resolved stars in 100 star 
clusters thus derived. Some CMDs include only a few resolved stars. 
Nevertheless they show features that can be used for age estimation 
such as bright MS in young star clusters. 
The most reliable age determination of star cluster can be made with the 
location of the MSTO. We regarded the position of the brightest MS star 
as MSTO for young star clusters. However, even the deepest image set 
among these could not reveal the positions of the MSTO for old star clusters. 
In those cases, we had to rely on the RGB feature which is not sensitive 
to the age. Therefore, we estimated only the lower  age limits of 
the star clusters by fitting the isochrones to the observed part of the
RGB.

We used the theoretical isochrones in the Padova models \citep{gir02}  to estimate the age
of each star cluster. 
\citet{cha99b} reported that metallicities of their 60 star clusters
span a large range from $Z = 0.0002$ to $Z = 0.03$, and that their young cluster samples
can be well represented by the $Z=0.004$ model, while there is a distinct turnoff
to lower metallicity($Z=0.001$) around $log(t)\sim 9.0$. 
Accordingly, we used the model for $Z=0.004$
for star clusters that turned out to have mainly blue or intermediate color stars and few 
red stars.
As for the old star clusters whose detected members are mainly RGB stars, we used
either model of $Z=0.004$ or $Z=0.001$, whichever metallicity could fit the RGB better. 
We used the reddening values given in the \citet{par07} for isochrone fitting for each 
star cluster. 

We adopted as the distance to M33, 910 kpc ($(m-M)_0 = 24.8$) given by
\citet{kim02}. Note there is a good summary of recent estimates of the distance to 
M33 obtained in various ways given by \citep{gal04}.
\citet{kim02} estimated the distance to M33 using the $I$-band magnitude of the tip of red 
giant branch (TRGB) of field stars \citep{lee93}. 
During this study, we visually checked 
that the $I$-band magnitude of TRGB for M33 field stars in our images is indeed close to their 
estimation. Therefore we adopted their result as the distance to M33.

Isochrones were fitted visually, and the upper and lower age limits of
a star cluster were also estimated visually.
Table~\ref{cluster_age} lists the ages of the M33 star clusters we derived, along with
metallicities and the reddening values used.

\subsection{Age distribution}

Fig.~\ref{agedist} shows the age distribution for 100 clusters including the
ten star clusters with lower age limit only, 
in comparison with those in previous studies:
(a) Chandar sample for 103 star clusters from  \citet{cha99b, cha02} who used
integrated photometry for age estimation, (b) Ma sample for 226 star clusters from 
\citet{ma01, ma02a, ma02b, ma02c, ma04a, ma04b} who also used integrated photometry 
for age estimation, and (c) Sarajedini sample for 20 star clusters from \citet{sar07a}
who used the CMD of the resolved stars in clusters for age estimation. 

The age distribution of star clusters in this study shows a broad
distribution with a peak at $log(t) \sim 7.8$, and a small number of
star clusters older than $log(t) \sim 9.0$.
The broad component can be fit well with a Gaussian 
with the peak age at $log(t) = 7.83\pm0.03$ and the width of $\sigma=1.17\pm0.11$.
Thus a majority of the clusters in this study are
young to intermediate-age and a small number of the clusters are older than 
about one Gyr. 
It is  noted that there is only one star cluster younger than $log(t) = 7 $ in the our sample, 
while the Ma sample shows a large component of young clusters with a peak at
$log(t) =7$ and the Chandar sample shows a small component of young clusters 
with a peak at $log(t) =6.7$.

Due to the different depths of HST/WFPC2 images used in this study, it is not easy 
to estimate the completeness of the age sample. 
However, we note that the isochrone fitting method used in this study has a bias such 
that bright, large sparse star clusters are likely to be included more than faint
compact ones in the final age sample.

\subsection{Integrated color-magnitude diagram} 

\begin{figure}
\epsscale{1.00}
\plotone{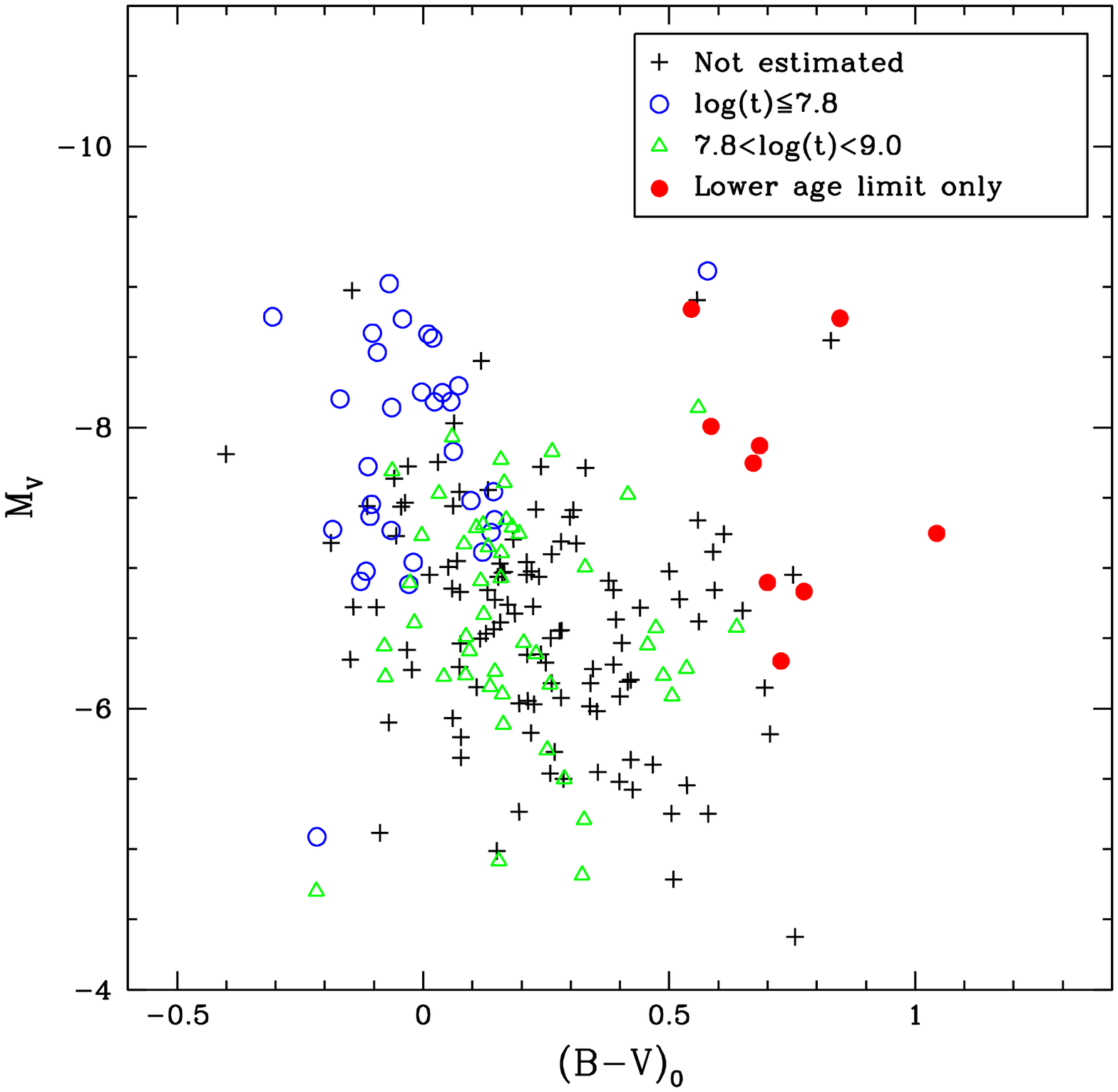}
\caption{
Integrated $M_V - (B-V)_0$ CMD of 242 star clusters 
in the sample of \citet{par07}. 
Open circles, open triangles, and filled circles represent, respectively,
young clusters ($\log (t) \leq 7.8$), intermediate-age clusters ($7.8<\log (t)<9$), 
and old star clusters with lower age limits of $\log (t) \sim 9$. 
Pluses  represent the clusters with no age estimates.
\label{intcmd}}
\end{figure}

Fig.~\ref{intcmd} displays the CMD of integrated photometry for 242 star clusters 
in M33 
from \citet{par07}, along with the information of age derived in this study. 
In general,
the age and integrated magnitude/color show consistent behaviors on the CMD.
Old star clusters  with  $log(t) > 9.0$ have colors $-0.5<(B-V)_0<1.1$ and magnitudes 
$-6<M_V<-9$, which are similar to those 
of the  Milky Way globular clusters (MWGCs), $0.6 \leq (B-V)_0 \leq 1.0$ and $-10.0 
\leq M_V \leq -6.0$. 
Therefore, these old star clusters can be regarded as being analogous to the MWGCs.

One notable star cluster is cluster ID 85 (= MKKSS-27, SM-261), 
with $M_V = -9.114$ and $(B-V)_0 = 0.578$.
It is the brightest in the sample.
These values are typical magnitude and color of an old globular cluster. 
\citet{cha02} estimated its age to be $log(t) = 8.6\pm 0.2$
from integrated photometry, or 2-5 Gyrs from spectroscopy. 
We obtained PSF photometry of ten stars around this star cluster. 
Summing the magnitudes of these ten stars, we derive the total magnitude and color
$V = 16.051\pm0.002$ and $(B-V) = 0.657\pm0.004$. 
These values are very similar to those from aperture photometry,
$V = 16.017\pm0.006$ and $(B-V) = 0.678\pm0.004$ \citep{par07}.
Therefore we regarded PSF
photometry recovered most of bright member stars in this cluster.
The HST/WFPC2 images of the field including this cluster and the CMD of this cluster 
show that there are two very bright red stars.
The spatial locations of these 
red stars in the cluster indicate that they are probably member stars, 
and their locations on the CMD show that they are red supergiant stars.
The fluxes of these stars contribute to most of the integrated light of the
star cluster.
If we remove the light contribution of these two red supergiant stars, 
we get the integrated photometry of $V=18.625\pm0.015$, and $(B-V)=-0.004\pm0.019$.
This shows that this cluster is much younger than the age derived from the case including
the two stars.
We estimated the age of this star cluster to be $log(t)=7.0$, which is much
younger than previous estimations. 
This case clearly shows the power of age estimation with resolved star CMD over those based on
integrated light.

Intermediate-age star clusters ($7.8 < log(t) < 9.0$) occupy the intermediate color
region in the CMD. It is interesting to note that their integrated magnitude range is quite 
fainter than those of younger star clusters or of older red star clusters. 
On the other hand, star clusters  younger than 
$log(t)\sim8.0$ have bright young stars whose fluxes contribute the majority of 
total star cluster light. These stars are bright main sequence stars or supergiants.
Consequently, young star clusters are located in the bright, blue regions on integrated
CMD.

\begin{figure}
\epsscale{1.00}
\plotone{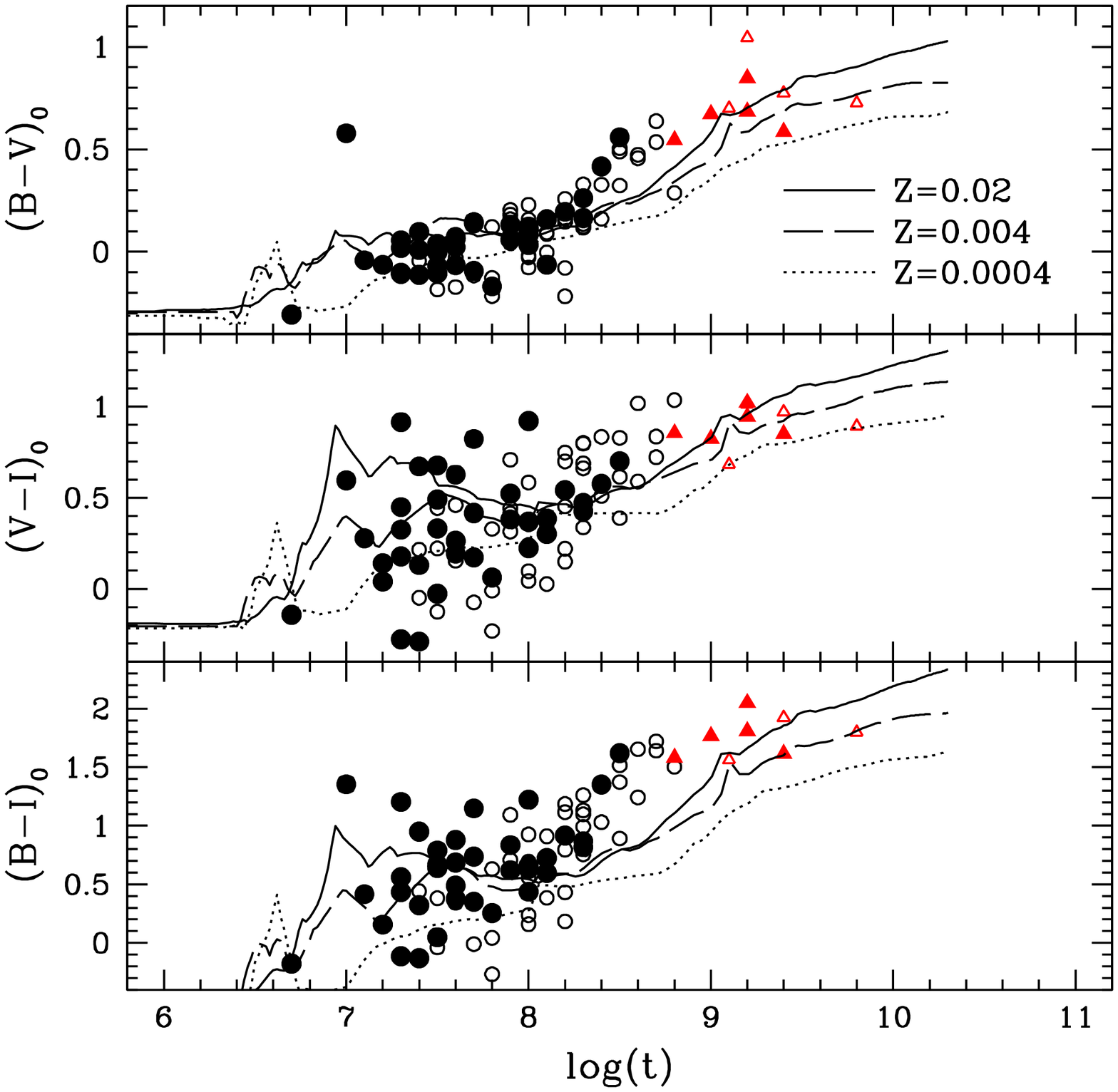}
\caption{
Integrated colors versus ages of the M33 star clusters. 
Also drawn are the theoretical curves predicted from single 
stellar population models of \citet{bru03}: for $Z=0.02$  (solid lines),
$Z=0.004$ (dashed lines) and  $Z=0.0004$ (dotted lines).
Circles denote the star clusters that have their age estimates as well as the
upper and lower age limits, 
while triangles denote those with lower age limits only. 
Filled and open symbols denote the bright star clusters ($V\leq18.0$ mag),
and faint clusters ($V > 18.0$ mag), respectively. 
\label{agecolor}}
\end{figure}

Since the ages in this study were derived in an independent way from using the integrated
color of the star clusters, more insight to star cluster properties can be obtained by 
comparing the ages and the integrated colors of the star clusters. 
Fig.~\ref{agecolor} displays the integrated color versus age diagram.
We plotted also 
the theoretical relations derived from single 
stellar population models of \citet{bru03}, for $Z=0.02$, 0.004 and 0.0004.
The figure shows that the integrated colors of the M33 clusters increase slowly as 
the age increases, approximately consistent with the theoretical curves.
$(V-I)$ and $(B-I)$ colors of the young clusters show a  large scatter for young 
star clusters, while their $(B-V)$ colors show a small scatter around the theoretical curves.
It is suspected that this may be due to the large uncertainty in the $I$-band photometry 
of the clusters. It is needed to obtain better integrated photometry in the $I$-band.

\subsection{Radial distribution\label{radial_dist}}

\begin{figure}
\epsscale{1.00}
\plotone{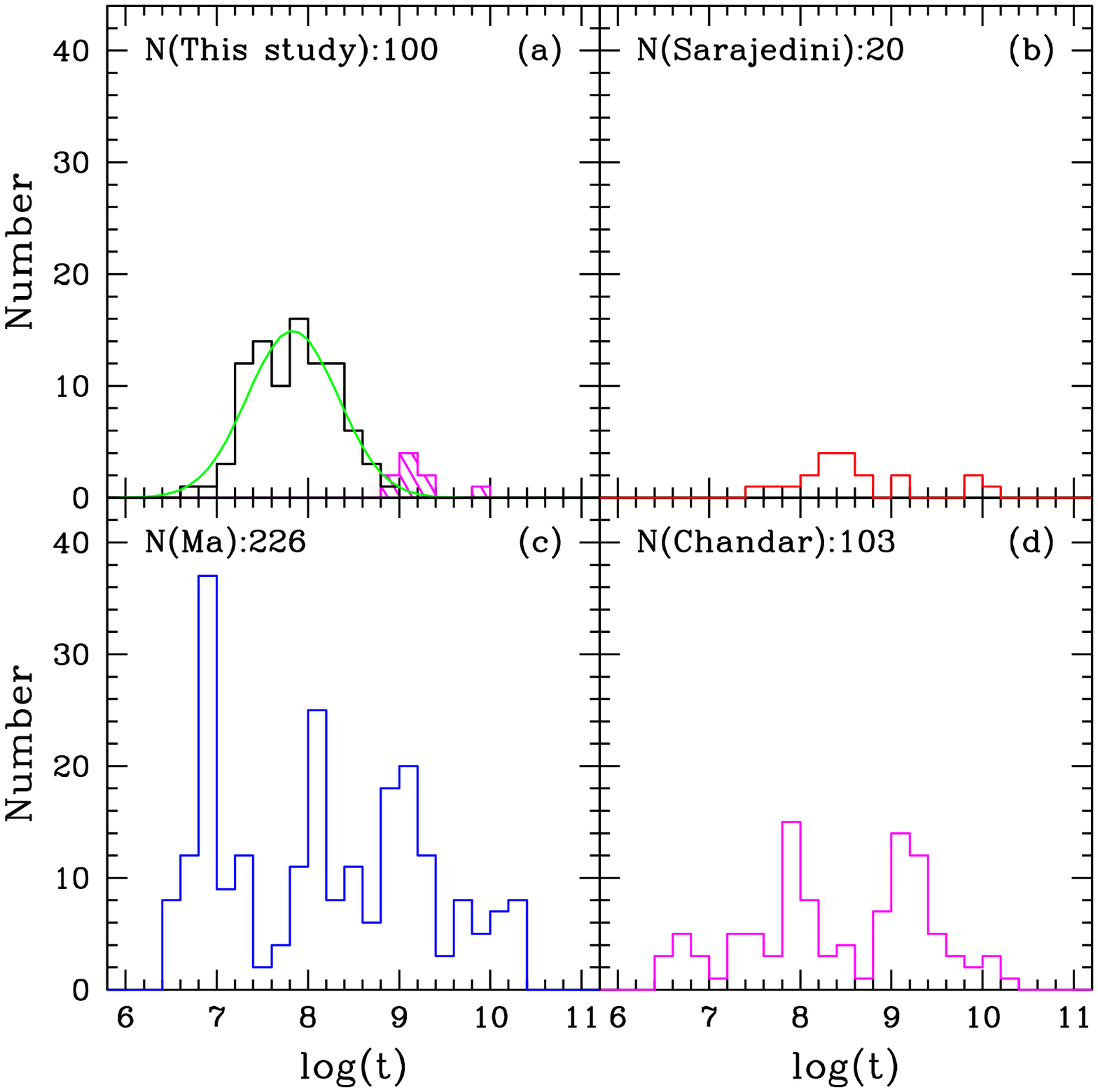} 
\caption{ 
Age distribution of M33 star clusters from four samples based on 
independent studies: (a) 100 star clusters from this study, 
(b) 20 star clusters from \citet{sar07a},
(c) 226 star clusters from \citet{ma01, ma02a, ma02b, ma02c, ma04a, ma04b}, and
(d) 103 star clusters from  \citet{cha99b, cha02}. The hatched histogram in panel (a) 
shows the distribution  for star clusters with lower age limits only in this study.
\label{agedist}}
\end{figure}

With 100 star cluster samples of known ages in hand, we investigated their radial 
distribution  around M33. 
With consideration of their integrated $(B-V)_0$ colors and the age distribution 
shown in Fig.~\ref{agedist}, we divided the star clusters into three age groups:
(1) young  star  clusters with $log(t) \leq 7.8$,
(2) intermediate-age star clusters with $7.8 < log(t) < 9.0$, and 
(3) old star clusters with $log(t) \geq 9.0$. 
The boundary value of $log(t) = 7.8$ divides the young population in half seen in Fig.~\ref{agedist}, 
and this age corresponds  to the color $(B-V)_0\sim 0.2$.

\begin{figure}
\epsscale{1.00}
\plotone{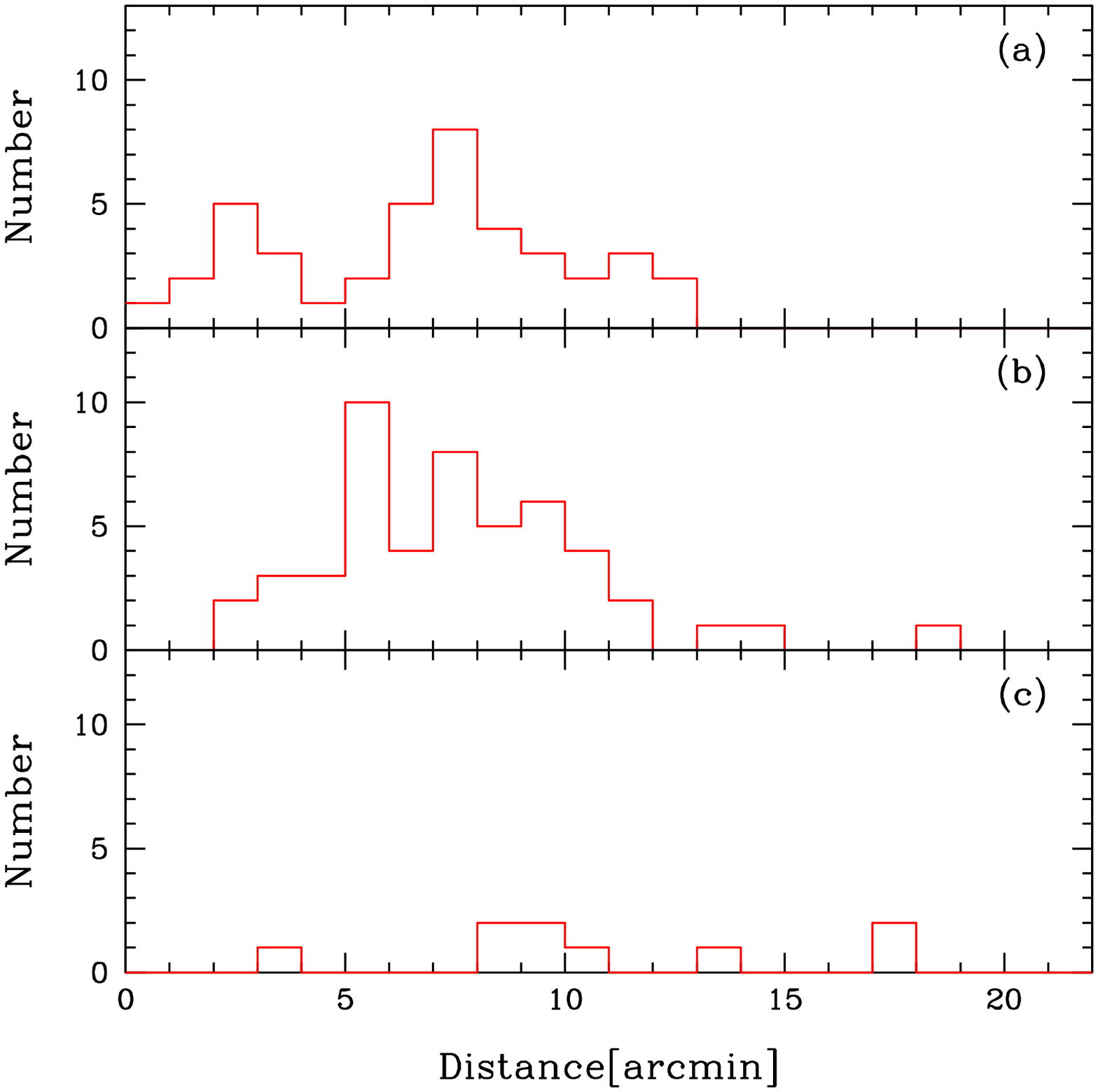}
\caption{
Radial distribution of the M33 star clusters:
(a) young clusters ($\log (t) \leq 7.8$), (b) intermediate-age clusters ($7.8<\log (t)<9$), 
(c) old star clusters with lower age limits of $\log (t) \sim 9$.
\label{radialage}} 
\end{figure}

Fig.~\ref{radialage} shows the number of the clusters as a function of the projected 
radial distance from the center of M33 
($\alpha(2000) = $1:33:50.9, $\delta(2000)= $30:39:35.8, \citep{skr06}).
Although it is not easy to investigate with small number of samples,
there exists a tendency that older star clusters are distributed in 
a wider area than the young star clusters in M33. 
Only young star clusters are found in the region close to the M33 center,  
whereas only intermediate-age and old star clusters are found in the outside region. 
This is consistent with the finding  by \citet{sar07b}
who used the age estimates derived from integrated photometry.

\section{Discussions}

\subsection{Comparison with other studies\label{s_agecomp}}

\begin{figure}
\epsscale{1.00}
\plotone{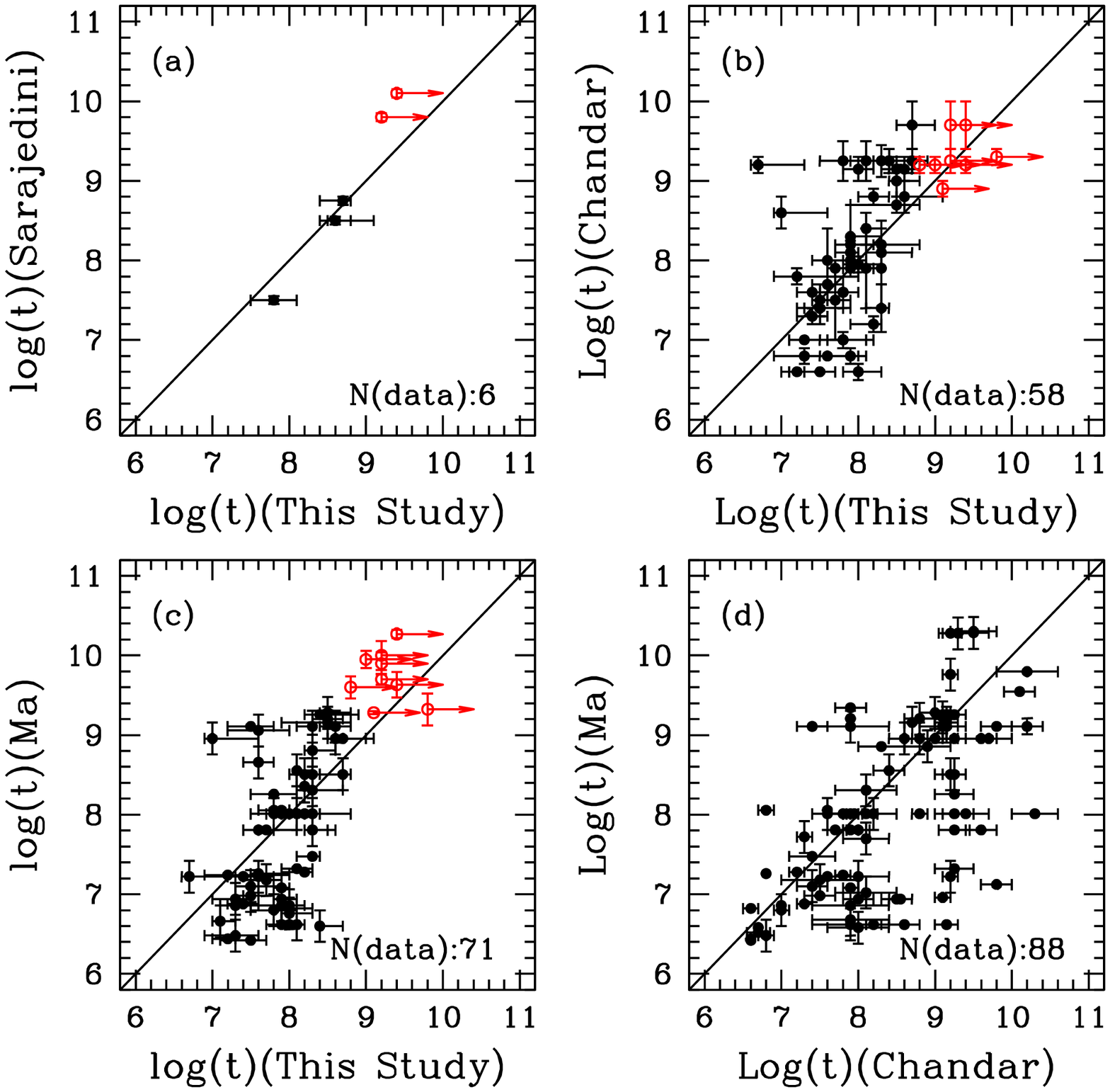} 
\caption{ 
Comparison of age estimates in this study with those from
(a) \citet{sar07a}, (b) \citet{cha99b, cha02},
and (c) \citet{ma01, ma02a, ma02b, ma02c, ma04a, ma04b}. Panel (d) shows the comparison
between \citet{cha99b, cha02} and \citet{ma01, ma02a, ma02b, ma02c, ma04a, ma04b}.
Open circles with arrows represent the star clusters with lower age limits.
\label{agecomp}}
\end{figure}

\citet{sar07a} estimated the ages of 20 star clusters by fitting isochrones to
the stellar photometry data derived from HST/ACS images.
\citet{sar07a}'s photometry is 
deep enough to reach the MSTO for most of their target star clusters.
There are six star clusters with age estimates common between this study and \citet{sar07a}'s:
Clusters 
ID 120 (CBF-110, SM-103), 130 (CBF-108, SM-127), 131(CBF-106, SM-130), 
228 (CBF-70, SM-402), 231 (CBF-79, SM-409), and 232 (CBF-78, SM-410).
Fig.~\ref{agecomp} displays the comparison of two age estimates for these star clusters.
The age estimates show a good agreement with \citet{sar07a}'s for four clusters.
We derived only the lower age limits for two clusters: $log(t) \geq 9.4$ for ID 228 and
$log(t) \geq 9.2$ for ID 131, while \citet{sar07a} derived $log(t) = 10.1 \pm 0.05$ and
$9.8 \pm0.05 $, respectively.

Age estimation of M33 star clusters with integrated photometry was carried out 
by \citet{cha99b, cha02} for most of their 154 star cluster samples. \citet{cha02} obtained
ages for additional 14 star clusters from the catalog of \citet{moc98} with the same method as in
\citet{cha99b}, in addition to revising the age estimates 
for some clusters with spectroscopy. 
There are 58 star clusters with age estimates common between theirs and this study.
Fig. \ref{agecomp}(b) shows that both age estimations 
agree within the error of measurements for many of star clusters. But our age estimates
for star clusters of $log(t) = 8.0\sim 9.0$ are younger than theirs in general.
On the other hand, there are considerable differences for young clusters such that 
our estimates are mostly $log(t) \geq 7.0$, while theirs are $log(t) \leq 7.0$.

\begin{figure}
\epsscale{1.00}
\plotone{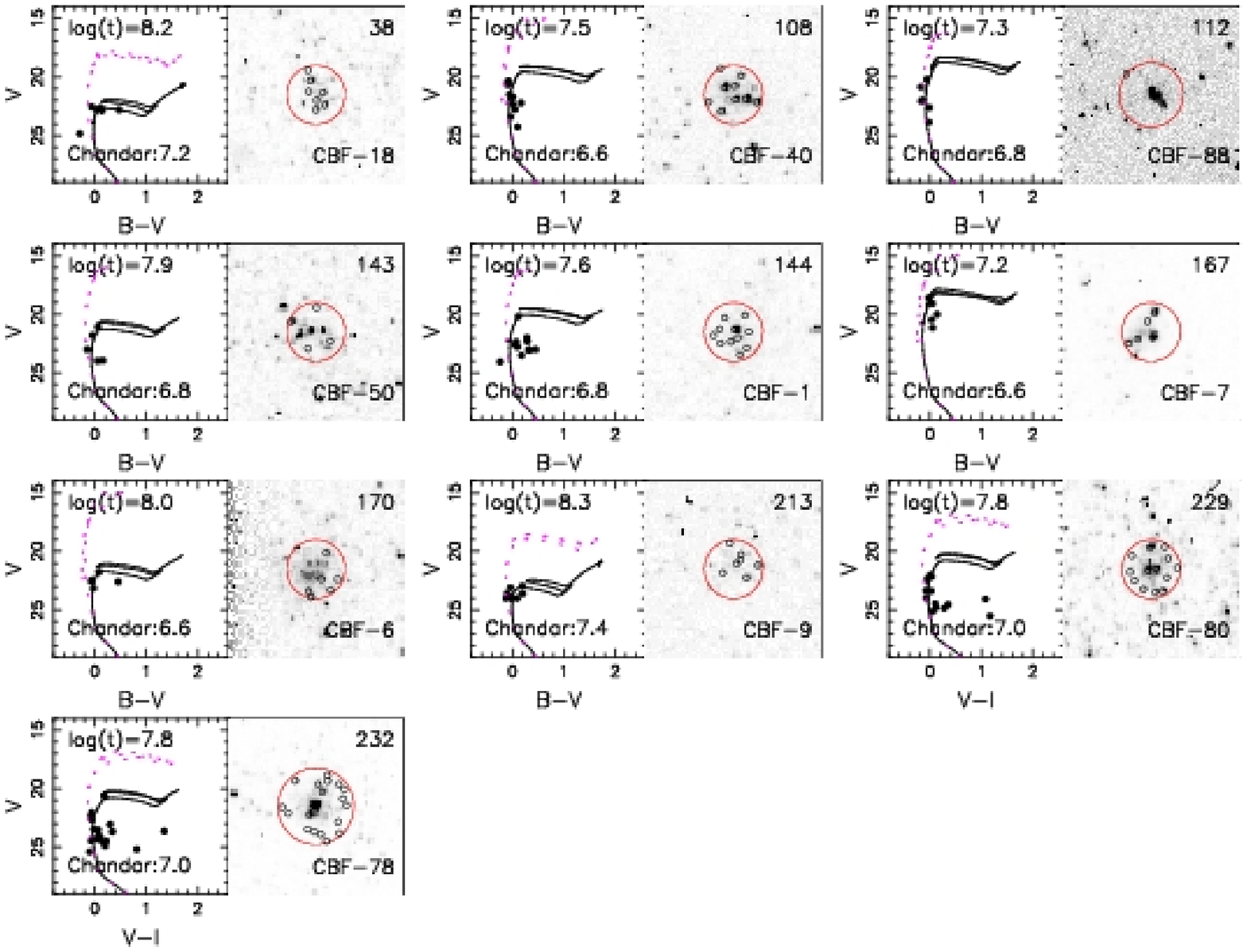}
\caption{ 
CMDs and $V$-band grayscale images of the star clusters for which our age estimates 
are larger than those by \citet{cha02}.
Intensity scales for grayscale maps are intentionally stretched so that the bright sources
in each cluster can be clearly distinguished at the expense of faint sources. 
A large circle in each grayscale image denotes the member selection
radius listed in Table~\ref{cluster_age}.
Solid lines and dashed lines in the CMDs represent the Padova isochrones for our age 
estimates and \citet{cha02}'s, respectively.
The points in each CMD represent the stars  marked with open circles on the 
corresponding grayscale image. 
\label{young_bigdiff}}
\end{figure}

We carefully examined the star clusters for which our age estimates are much
different from \citet{cha99b, cha02}. 
Fig.~\ref{young_bigdiff} shows the CMDs of ten young star clusters that 
have large age differences.
To compare the age estimations, we plotted the Padova isochrones for both age
estimates on each CMD. 
$V$-band grayscale images of the star clusters are also
displayed to show the spatial distribution of detected stars. 
It is seen that there are some very bright stars that were not detected  by HSTphot 
in clusters ID 143 and 232 (SM-168 and SM-410, respectively).
They might have led to younger age estimation of star clusters had they 
been detected and measured.
There are some undetected stars in cluster 
ID 170 (SM-220), but they look like faint stars. Except for above three star clusters, 
it seems that HSTphot detected all bright stars in the region of each star cluster. 
Their resolved star CMDs are more consistent with isochrones for our age 
estimates than those for their estimates. 
In short, seven out of ten star clusters show features that are more consistent 
with our age estimations than with theirs. 
Therefore, we argue that our age estimations are more reliable than
those by \citet{cha02} for most of star clusters with a large discrepancies in the age 
estimations.

\begin{figure}
\epsscale{1.00}
\plotone{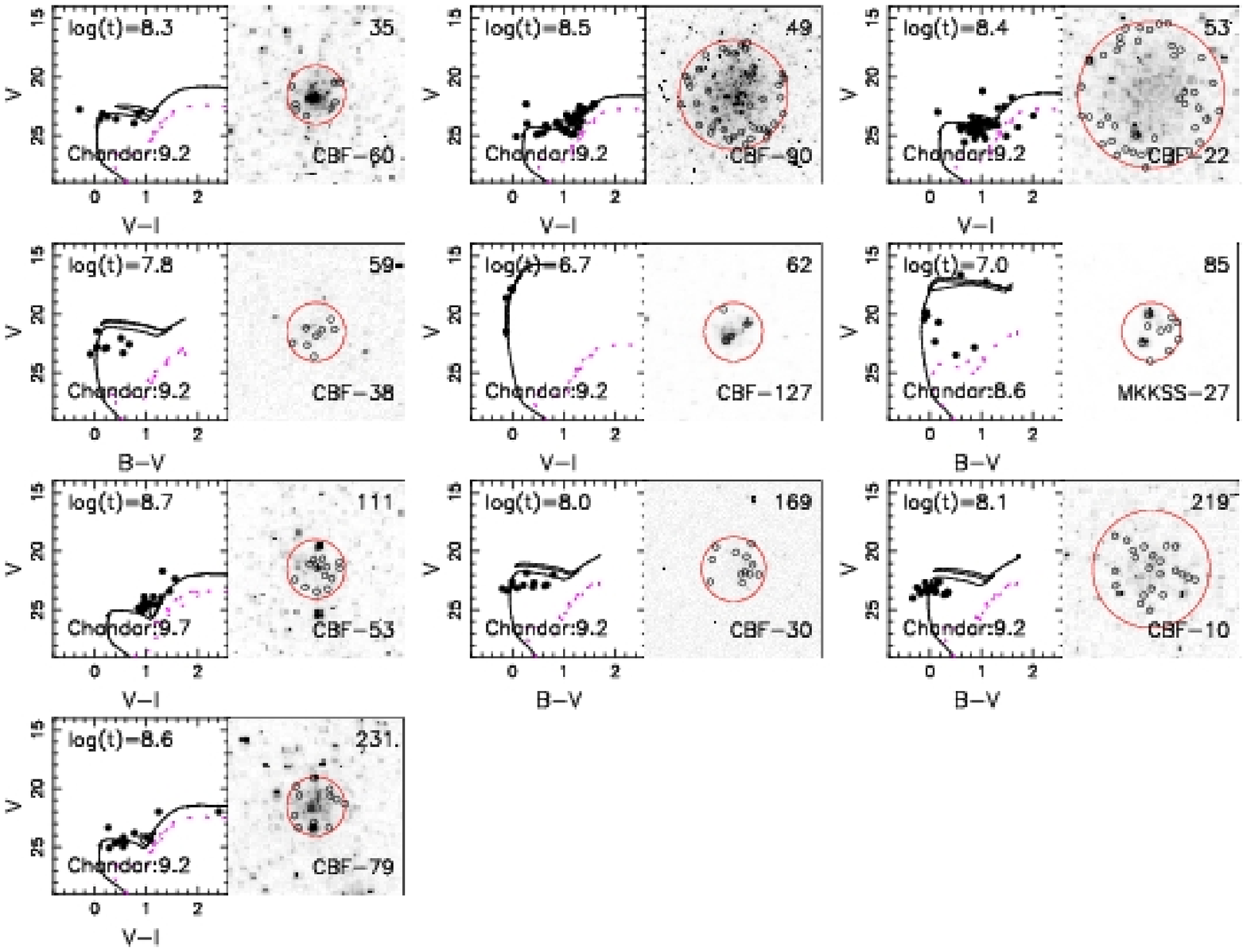}
\caption{ 
Same as Fig.~\ref{young_bigdiff}, but for star clusters that were estimated to be 
younger by us than by \citet{cha02}.
\label{old_bigdiff}}
\end{figure}

There is also considerable amount of age discrepancies for some older star
clusters: our age estimates are  $log(t) \leq 9.0$, 
while \citet{cha02} estimated their ages to be much older. 
The CMDs and $V$ grayscale images of these star clusters are shown in
Fig.~\ref{old_bigdiff}. 
Clusters ID 49, 53, 111, and 231 (SM-45, SM-95, SM-110, and SM-409, respectively).
turn out to have only red
star sequences in their resolved star CMDs. It is difficult to derive reliable
ages for these clusters just with RGB features. 

But for other star clusters, we noted that
there are some blue stars whose colors ($(V-I)$ or $(B-V)$) are $\sim 0.0$, with their
magnitude errors smaller than 0.2. Most of these stars are located in the very central
region of the star clusters. Therefore, we regard these as blue main sequence member stars.
If they are indeed the member stars of the cluster, 
\citet{cha02} might have overestimated the ages of
these star clusters. 
Note that \citet{cha99b} estimated from the integrated photometry
the ages($log(t)$) of clusters ID 35, 59, 169 and 219 (SM-68, SM-146, SM-219, 
and SM-377, respectively) 
to be $8.3\pm0.30$, $7.7\pm0.20$, $8.4\pm0.20$, and $7.8\pm0.30$, respectively,
which are very close to our estimates: $8.3\pm0.2$, $7.8\pm0.3$, $8.0\pm0.2$, and 
$8.1\pm0.2$.

We also compared our age estimates of the star clusters 
with the results obtained in a series of papers by
\citet{ma01, ma02a, ma02b, ma02c, ma04a, ma04b}, as shown in Fig.~\ref{agecomp}(c).
There are 71 star clusters common between 
both studies.
For star clusters older than $log(t) \sim 8.0$, 
our age estimates are in general younger than their estimates except for a few
star clusters. As for the star clusters of $log(t) \leq 8.0$, it turned out that our
age estimates are much older than their estimates, basically the same tendency as those
with \citet{cha99b, cha02} with larger differences.
We note they obtained $(B-V)$ colors to be bluer than \citet{cha99a} for the 
same clusters \citep{ma01}.
It seems that large seeing condition ($\sim 2.0''$) and large aperture ($r = 6.8''$)
used in their photometry caused light contamination from neighboring blue disk stars 
in M33, thus leading to younger age estimation even compared with those of \citeauthor*{cha99b}.
This can be seen in Fig.~\ref{agecomp}(d) where age estimates by \citeauthor*{ma01} and
\citeauthor*{cha99b} are compared. Although many of age estimates are in good agreement
between them over a wide age range, quite a number of star clusters show large differences,
For almost all of such star clusters, \citeauthor*{cha99b}'s estimates are
older than those by \citeauthor*{ma01}.

\citet{sar07b} compiled the data in the eight previous catalogs of M33 star cluster and
built a new catalog of  451 star
cluster candidates in M33, of which 255 are star clusters confirmed  with HST images. 
Since their age estimation is based on the results by \citeauthor*{ma01}, we do not show
the result here but refer to Fig.~\ref{agecomp}(c) for the age comparison between ours 
and theirs.

These comparisons indicate that the estimated ages for some star clusters in this study have
uncertainties due to the possible inclusion of non-member M33 field stars. This is true to
both integrated photometry or PSF photometry of resolved stars.
However, the effect of non member star selection can be reduced in resolved star
photometry since the contribution of individual stars can be evaluated. For example,
CMD of stars near star cluster center, and those of control field are compared to
minimize the field star contamination.
Although PSF photometry could not detect some bright stars in the cluster region, we  
argue that our age estimations for star clusters based on the isochrone
fitting are more accurate than other age estimations
based on the integrated light of the star clusters.

The resultant age distribution shows that there is only one star cluster which is younger
than $log(t)\sim 7.0$, as shown in Fig.~\ref{agedist}(a). 
It might be that there are indeed few young star clusters in M33.
However, M33 is a late spiral galaxy, in which signatures
for active star formation are clearly seen in a number of giant HII regions such as NGC 604 or
NGC 595 on its disk. Far infrared observation indicates M33 is more actively forming stars 
(with star formation rate $\sim 0.3-0.7 M_{\odot}$/year) than either our Galaxy 
or M31 \citep{hip03}.
Therefore it is unlikely that M33 lacks very young star clusters.
In fact, the presence of very young clusters that are 3--5 Myrs old was reported in 
NGC 604 \citep{hun96}. Such young clusters normally are embedded 
in nebulous environments, which makes it difficult to be detected with visual investigation.

\subsection{Formation of old star clusters}

The interaction between galaxies is known to often trigger the formation of stars
and star clusters. 
For example, 
a large population of young bright star clusters in the interacting galaxy pair of NGC 4038/4039 (Antennae galaxies) are believed to be involved with the interaction of their host galaxies \citep{whi99} .
\citet{leh05} suggested that the intermediate-color clusters in the 
interacting galaxy NGC 1487 were probably formed during the merging process. 
\citet{lee05}  found out that the age distribution of the star clusters in NGC 5194 shows a broad peak that 
is consistent with the crossing times of the companion galaxy NGC 5195 through NGC 5194
disk. These results indicate that the age distribution is consistent with the scenario that the interaction
between two galaxies directly caused an increase in cluster formation rate.

In the case of M33, the spatial distribution of halo stars shows no sign of disturbance \citep{fer06}. 
However, HI surveys around M33 revealed an HI bridge between M31 and M33 \citep{bra04}, 
and the outer HI gas disk of M33 is found to be skewed towards the M31 direction \citep{cor97}.  
From the model simulation, 
\citet{bek08} suggested the tidal interaction between M31 and M33 about 4-8 Gyrs ago 
can explain the HI bridge and the observed HI warp in M33.
There are ten old star clusters with lower age limits  of $log(t) \sim 9.0$  in our sample.
While we cannot estimate their exact ages with out data, previous studies on 
some of these clusters suggested that they are several Gyrs younger than Galactic 
globular clusters (see \citet{sar00, cha06, zlo09}).
If the peak for the age distribution of these old star clusters are found to be
similar to the epoch of the interaction, it might be possible  
that the galaxy interaction with M31 had triggered the formation of old star 
clusters in M33.
We note that \citet{hux09} suggested the possibility that the tidal interaction 
between M31 and M33 have affected the spatial distribution of M33 globular clusters
at large galactocentric distances.

However, the small number of old star cluster samples prevents any definite conclusion on the picture of
their formation. Further studies with more sample with good age estimation are 
needed to investigate in detail the formation of M33 star cluster system. 

\section{Summary and Conclusion}

We present $BVI$ photometry of the resolved stars in star clusters on the
HST/WFPC2 images of 79 fields in M33. 
Most star clusters in M33 are resolved into individual stars on the HST/WFPC2 images.
We estimate the ages of 100 star clusters matching the CMDs of resolved stars 
with Padova theoretical isochrones.

The age estimates for the star clusters in this study show a broad distribution
from $log(t) \sim 7.0$ to $log(t) \sim 9.0$ with a peak at $log(t) = 7.82 \pm 0.03$.
There are also a small number old star clusters with  lower age limits of $log(t) \sim 9.0$. 
It is found out that there  are few clusters younger than $log(t) = 7.0$ in this study.

Previous estimates \citep{cha99b, cha02, ma01, ma02a, ma02b, ma02c, ma04a, 
ma04b} for some young star clusters show a difference from our estimates.
Close inspection of the CMDs and images of each star cluster reveals that previous estimations
with integrated light underestimated the ages for about 80\% of young star clusters that have
significant differences among the age estimations.

Age distribution along the galactocentric distance reveals that there is a hint that blue young star 
clusters show a stronger concentration toward the M33 center than older star clusters.
The interaction between M31 and M33 might affect the formation and
spatial distribution of old star clusters in M33.

\acknowledgments
The authors thank anonymous referee for helpful discussions and comments.
M.G.L. is supported 
in part by a grant
(R01-2007-000-20336-0) from the Basic Research Program of the
Korea Science and Engineering Foundation.
W. -K. P. acknowledges the supports from Astrophysical Research Center for the Structure 
and Evolution of the Cosmos (ARCSEC), at Sejong University, Korea, and from Korea Institute 
for Advanced Studies (KIAS).

\clearpage

\clearpage

\clearpage

\begin{figure}
\figurenum{2}
\epsscale{0.95}
\plotone{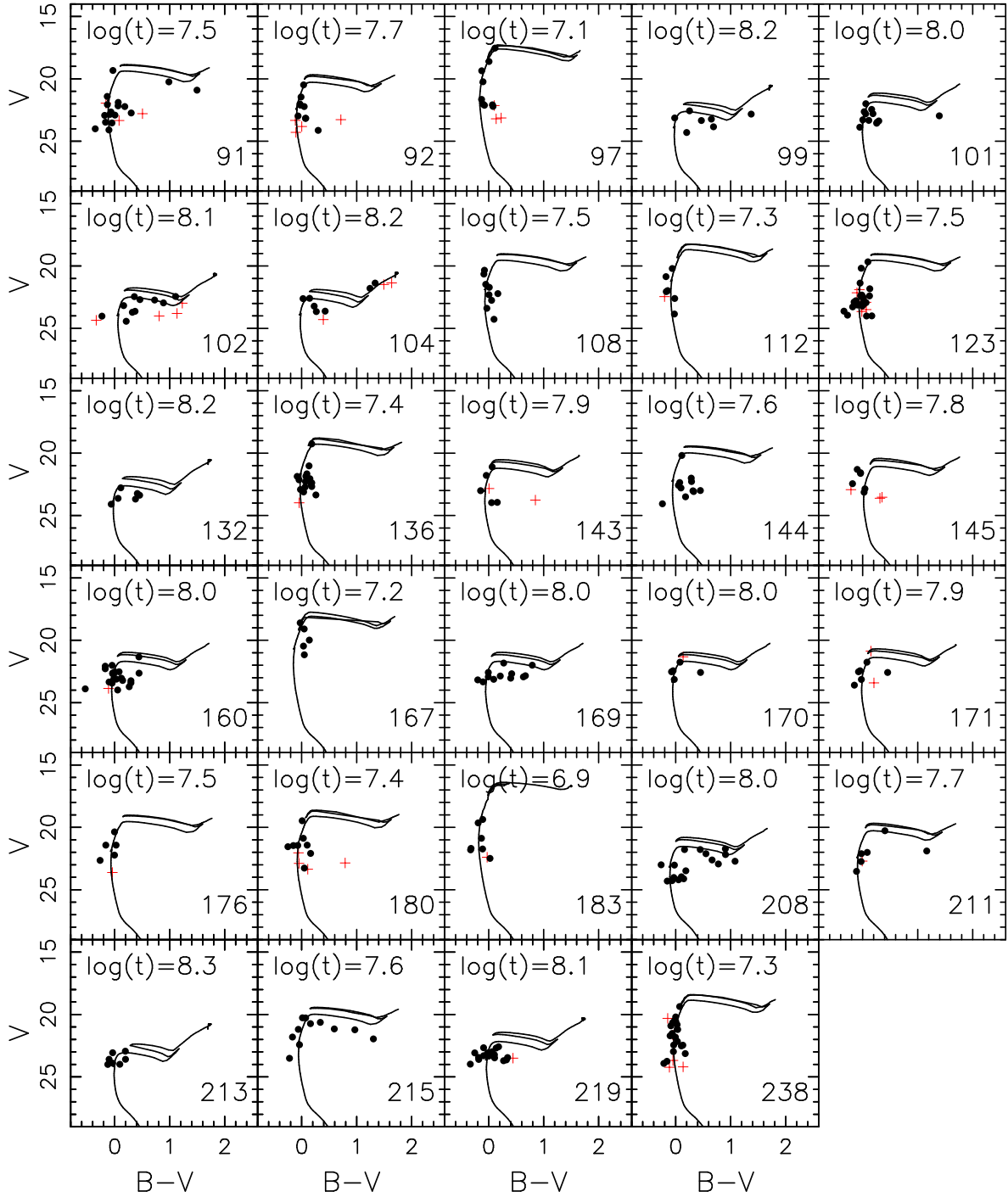} 
\caption{ 
$V-(B-V)$ CMDs continued.}
\end{figure}
\clearpage

\begin{figure}
\figurenum{2}
\epsscale{0.95}
\plotone{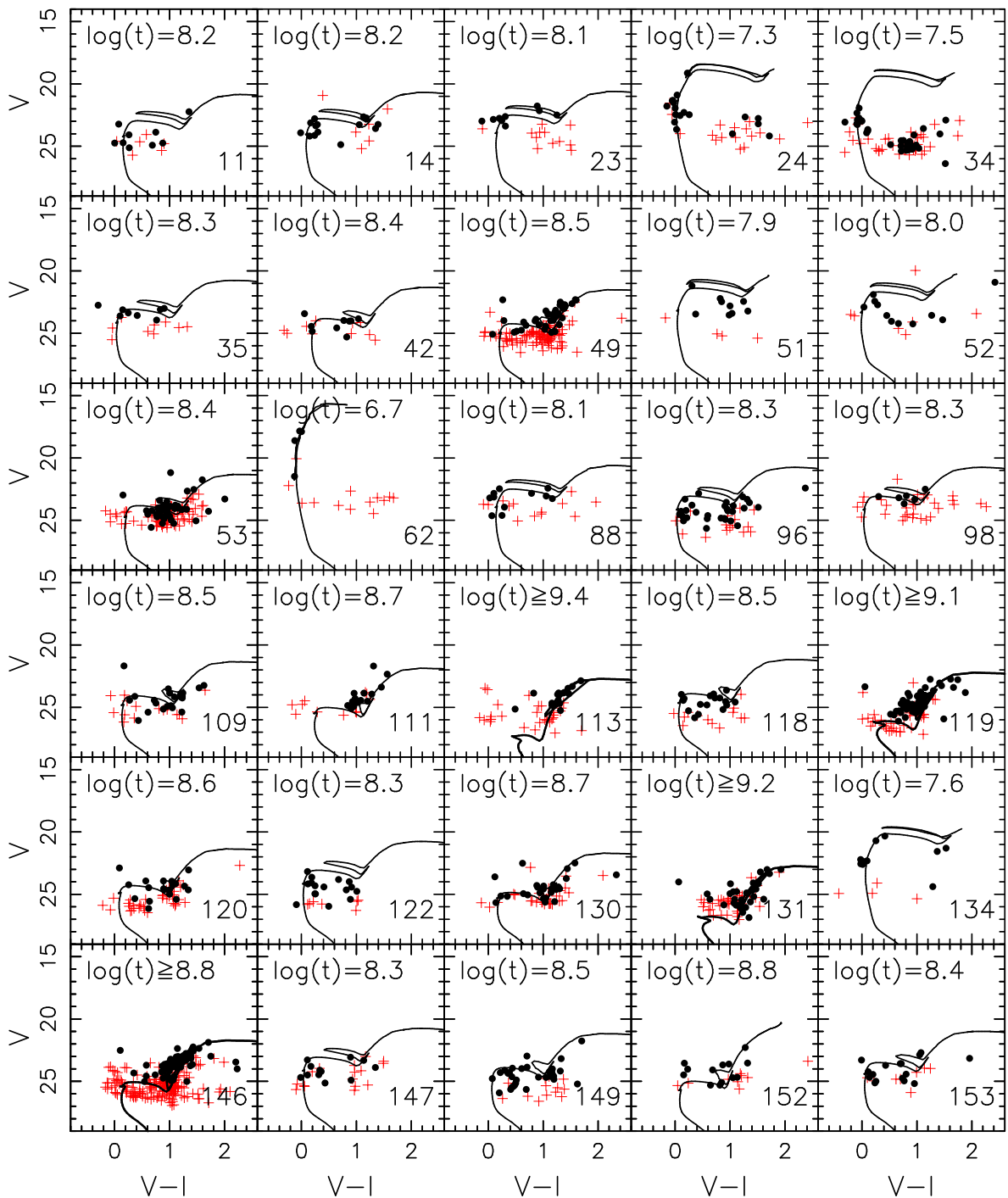} 
\caption{ 
$V-(V-I)$ CMDs of the resolved stars in M33 star clusters.}
\end{figure}
\clearpage

\begin{figure}
\figurenum{2}
\epsscale{0.95}
\plotone{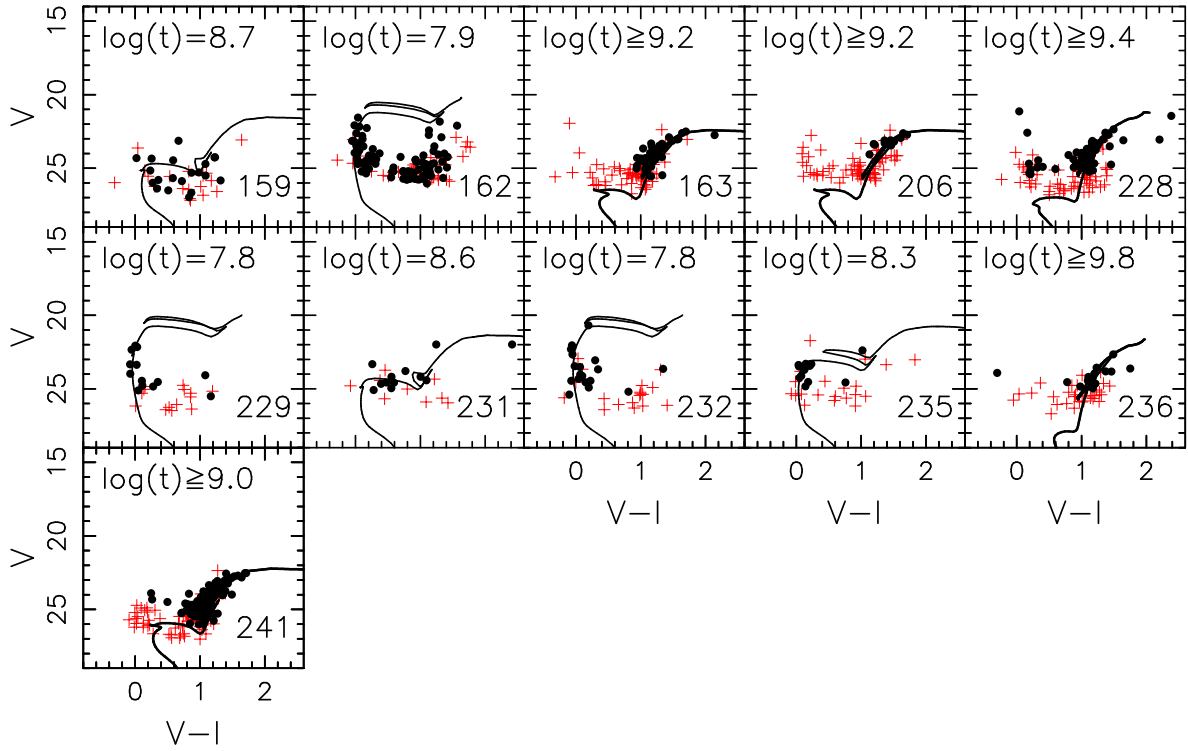} 
\caption{ 
$V-(V-I)$ CMDs continued.}
\end{figure}
\clearpage

\clearpage

\clearpage

\clearpage

\clearpage

\begin{deluxetable}{crrrcl}
\tabletypesize{\footnotesize}
\tablecaption{Age estimates for M33 star clusters\label{cluster_age}}
\tablewidth{0pt}
\tablehead{
\colhead{ID\tablenotemark{a}} & \colhead{${\rm log(t)}$} & \colhead{Z} &
\colhead{$E(B-V)$\tablenotemark{a}} & \colhead{r(arcsec)} & \colhead{Cross identification\tablenotemark{b}}
}
\startdata
101 & $ 8.0^{+0.1}_{-0.2} $ &   0.004 &         $ 0.15 \pm 0.05$ & $ 1.0 $ & CS-U70, SM-342 \\
102 & $ 8.1^{+0.2}_{-0.0} $ &   0.004 &         $ 0.20 \pm 0.05$ & $ 1.2 $ & CS-C33, MD-36, MKKSS-45, SM-351 \\
104 & $ 8.2^{+0.2}_{-0.2} $ &   0.004 &         $ 0.10 \pm 0.05$ & $ 1.0 $ & CBF-154, CS-H19, SM-355 \\
108 & $ 7.5^{+0.2}_{-0.4} $ &   0.004 &         $ 0.15 \pm 0.05$ & $ 1.0 $ & CBF-40, SM-35 \\
109 & $ 8.5^{+0.3}_{-0.6} $ &   0.004 &         $ 0.15 \pm 0.05$ & $ 1.5 $ & CBF-86, CS-U140, SM-40 \\
111 & $ 8.7^{+0.3}_{-0.2} $ &   0.004 &         $ 0.20 \pm 0.05$ & $ 1.0 $ & CBF-53, SM-45 \\
112 & $ 7.3^{+0.3}_{-0.4} $ &   0.004 &         $ 0.15 \pm 0.05$ & $ 1.0 $ & CBF-88, SM-47 \\
113 & $ \geq 9.4 $ &   0.004 &         $ 0.10 \pm 0.05$ & $ 1.2 $ & CBF-54, CS-U137, MD-8, SM-49 \\
118 & $ 8.5^{+0.2}_{-0.1} $ &   0.004 &         $ 0.10 \pm 0.05$ & $ 1.0 $ & CBF-111, SM-100 \\
119 & $ \geq 9.1 $ &   0.004 &         $ 0.10 \pm 0.05$ & $ 1.5 $ & CBF-114, CS-C38, MD-11, SM-102 \\
\enddata
\tablenotetext{a}{Taken from \citet{par07}}
\tablenotetext{b}{B identifications are from \citet{bed05}; CBF identifications from \citet{cha99a, cha01};
CS identification from \citet{cs82}; MKKSS identifications from \citet{moc98}; MD identifications are
from \citet{md78}; SBGHS identifications from \citet{sar07a}; and SM identifications from \citet{sar07b}}
\tablecomments{The complete version of this table is in the electronic edition of
the Journal. The printed edition contains only a sample.}
\end{deluxetable}


\end{document}